\documentclass[twocolumn,twoside]{revtex4}
\usepackage{amsmath}
\usepackage{amssymb}
\usepackage{graphicx}
\usepackage{subfigure}
\usepackage{hyperref}
\usepackage{fancyhdr}
\newcommand{\etc}{etc\ldots}
\newcommand{\ie}{i.e.,}

\newcommand{\lsr}{\mathcal{R}}
\newcommand{\qcd}{\text{QCD}}
\newcommand{\had}{\text{had}}
\newcommand{\double}[2]{(#1,\,#2)}
\newcommand{\stud}{\rule[-1ex]{0ex}{1ex}}

\begin{document}

\title{Meson-Hybrid Mixing in Vector ($1^{--}$) and Axial Vector ($1^{++}$) Charmonium}

%

\author{D.\ Harnett}
\affiliation{Department of Physics, University of the Fraser Valley, Abbotsford, BC, Canada, V2R 3E7}
\author{A.\ Palameta}
\author{J.\ Ho}
\author{T.\ G.\ Steele}
\affiliation{Department of Physics and Engineering Physics, University of Saskatchewan, Saskatoon, SK, Canada, S7N 5E2}

\begin{abstract}
\noindent We study mixing between conventional and hybrid mesons in vector and axial vector charmonium using QCD Laplace sum-rules. 
We compute meson-hybrid cross correlators within the operator product expansion, taking into account condensate contributions up to and 
including those of dimension-six as well as composite operator renormalization-induced diagrams. 
Using measured masses of charmonium-like states as input, we probe known resonances for nonzero coupling to both conventional and hybrid meson currents, 
a signal for meson-hybrid mixing. 
\end{abstract}

\maketitle

\thispagestyle{fancy}


\section{Introduction}\label{I}
Hybrid mesons are hadrons containing a constituent quark, antiquark, and gluon.
As they are colour singlets, they should be observable.
Despite decades of searching, they have yet to be conclusively identified in experiment.
Hadron mixing, the idea that observed hadrons might be superpositions of 
conventional (\ie\ quark-antiquark) mesons, hybrid mesons,
tetraquarks, \etc, could be hampering identification.

To explore this idea, we consider the XYZ resonances, a collection of 
charmonium-like states many of which are not readily interpretated as
conventional mesons.  (For a review, see~\cite{Brambilla}.)  
We focus on the vector (\ie\ $J^{PC}=1^{--}$) and axial vector (\ie\ $J^{PC}=1^{++}$)
channels.  Known resonances in these channels are listed in
Tables~\ref{vectorStates} and~\ref{axialStates}~\cite{PDG}.

\begin{table}[hbt]
\begin{center}
\caption{Known charmonium-like vector (\ie\ $J^{PC}=1^{--}$) resonances.}
\begin{tabular}{|l|c|}
  \hline
  \stud Name & Mass (GeV) \\
  \hline
  $J/\psi$ & 3.10 \\
  $\psi(2S)$ & 3.69 \\
  $\psi(3770)$ & 3.77 \\
  $\psi(4040)$ & 4.04 \\
  $\psi(4160)$ & 4.19 \\
  $X(4230)$ & 4.23 \\
  $X(4260)$ & 4.23 \\
  $X(4360)$ & 4.34 \\
  $\psi(4415)$ & 4.42 \\
  $X(4660)$ & 4.64 \\
  \hline
\end{tabular}
\label{vectorStates}
\end{center}
\end{table}

\begin{table}[hbt]
\begin{center}
\caption{Known charmonium-like axial vector (\ie\ $J^{PC}=1^{++}$) resonances.}
\begin{tabular}{|l|c|}
  \hline
  \stud Name & Mass (GeV)\\
  \hline
  $\chi_{c1}(1P)$  & 3.51\\
  $X(3872)$ & 3.87\\
  $X(4140)$ & 4.15\\
  $X(4274)$ & 4.27\\
  \hline
\end{tabular}
\label{axialStates}
\end{center}
\end{table}

We test the resonances of Tables~\ref{vectorStates} and~\ref{axialStates}
for coupling to a conventional meson-hybrid meson cross-correlator using 
QCD Laplace sum-rules (LSRs).  
QCD sum-rules are transformed dispersion relations
that relate a QCD-computed correlator to an integral over a hadronic 
spectral function~\cite{SVZ_I,SVZ_II}.  Using measured resonance masses 
(and effective widths) as input,
we extract products of conventional meson and hybrid meson couplings,
\ie\ mixing parameters, as best-fit parameters between QCD and hadron physics.
Resonances with nonzero mixing parameters can be interpreted as having both
conventional meson and hybrid meson components. 

\section{Correlators}\label{II}
\begin{figure}[hbt]
\centering
\includegraphics[width=0.9\columnwidth]{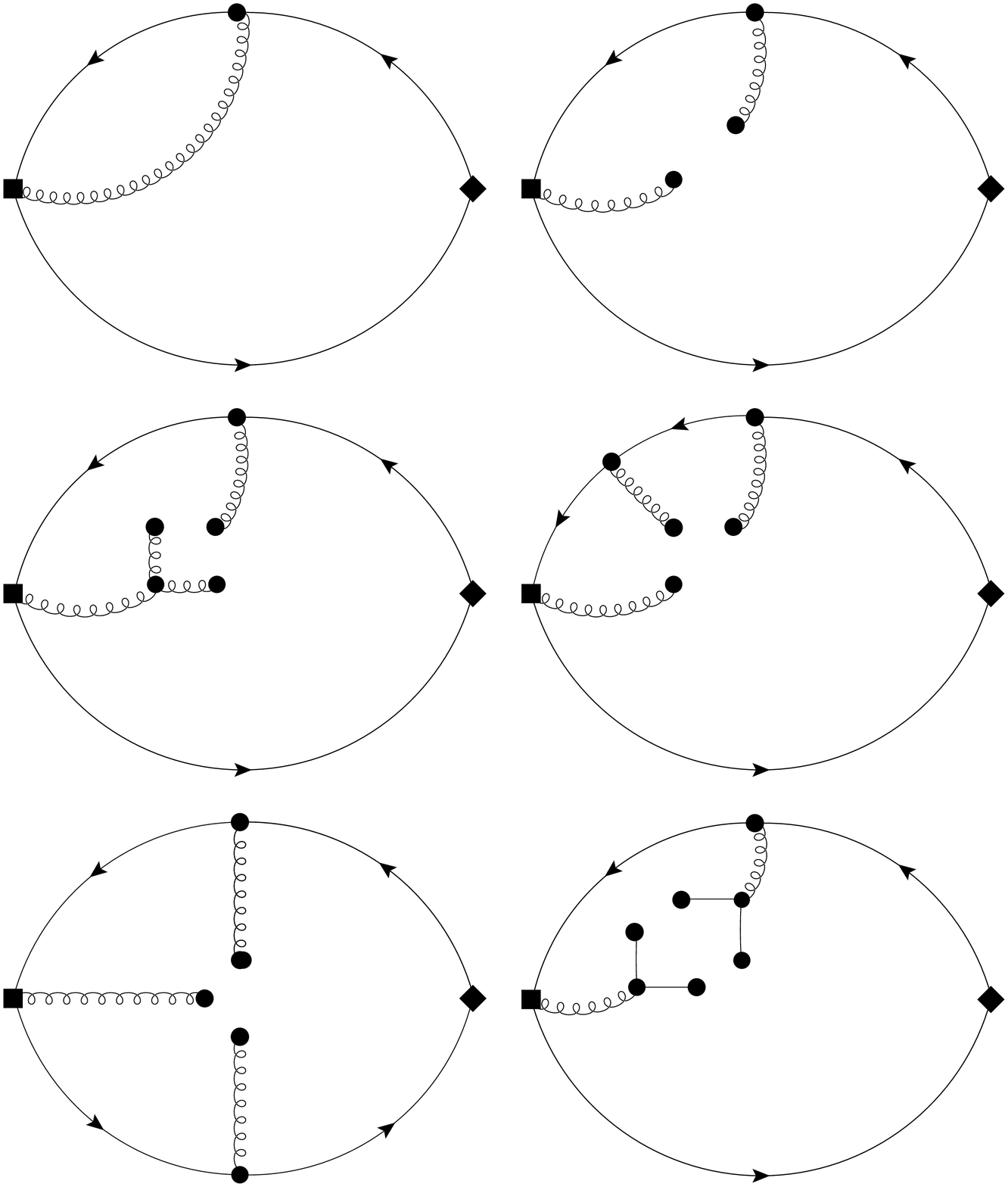}
\caption{\label{opeDiagrams} The diagrams calculated for $\Pi(Q^2)$ 
from~(\ref{correlator}). 
The diagram in the upper left is perturbation theory.  The others
are nonperturbative condensate contributions.
Square vertices represent the hybrid meson
current~(\ref{hybridCurrent}).  Diamond vertices represent the conventional 
meson current~(\ref{mesonCurrent}).
All Feynman diagrams in this manuscript were produced with 
JaxoDraw~\cite{jaxodraw}.}
\end{figure}

Consider the charmonium-like conventional meson-hybrid meson cross-correlator,
\begin{multline}\label{correlator}
\Pi(q^2) = \frac{i}{D-1}
  \left( \frac{q_{\mu}q_{\nu}}{q^2} - g_{\mu\nu} \right)\\
  \times\int\! d^D\! x\, e^{iq\cdot x}\,
  \langle\Omega|\tau j^{(m)}_{\mu}(x) j^{(h)}_{\nu}(0)|\Omega\rangle,
\end{multline}
for spacetime dimension $D$
between conventional meson current
\begin{equation}\label{mesonCurrent}
j^{(m)}_{\mu} = 
\begin{cases}
  \bar{c}\gamma_{\mu}c\ \text{for}\ 1^{--}\\
  \bar{c}\gamma_{\mu}\gamma_5 c\ \text{for}\ 1^{++}
\end{cases}
\end{equation}
and hybrid meson current
\begin{equation}\label{hybridCurrent}
j^{(h)}_{\nu} =
\begin{cases}
  g_s \overline{c} \gamma^{\rho}\gamma_5 \frac{\lambda^a}{2} 
 \left(\frac{1}{2}\epsilon_{\nu\rho\omega\eta}G^a_{\ \omega\eta}\right)c\ \text{for}\ 1^{--}\\
g_s \overline{c} \gamma^{\rho}\frac{\lambda^a}{2} 
 \left(\frac{1}{2}\epsilon_{\nu\rho\omega\eta}G^a_{\ \omega\eta}\right)c\ \text{for}\ 1^{++}
\end{cases}.
\end{equation}
In~(\ref{mesonCurrent}) and~(\ref{hybridCurrent}), $c$ is a charm quark,
$G^a_{\ \omega\eta}$ is the gluon field strength, and $\epsilon_{\nu\rho\omega\eta}$ is
the Levi-Civita symbol.
We compute $\Pi(Q^2)$ using the operator product expansion (OPE) in which 
perturbation theory is supplemented by nonperturbative corrections, each 
of which is the product of a perturbatively computed Wilson coefficient 
and a nonzero vacuum expectation value, \ie\ a condensate.
We consider condensates of dimension-six (\ie\ 6d) or less. 
Wilson coefficients are computed to leading-order in $\alpha_s = \frac{g_s}{4\pi}$.
The diagrams that contribute are shown in Fig.~\ref{opeDiagrams}.
Calculational details and correlator results can be found 
in~\cite{Palameta_I,Palameta_II}.  

\begin{figure}[hbt]
\centering
\includegraphics[width=0.9\columnwidth]{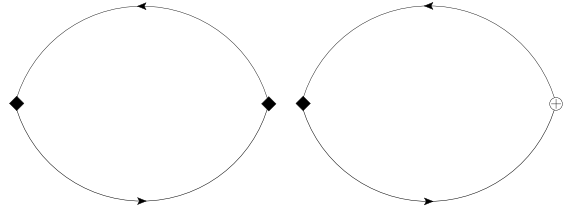}
\caption{\label{renormalizationDiagrams} The renormalization-induced diagrams 
that contribute to $\Pi(Q^2)$ from~(\ref{correlator}).
Diamond vertices represent the conventional meson current~(\ref{mesonCurrent}).
Thus circle-plus vertex corresponds to the covariant derivative current
$\bar{c}iD_{\nu}c$ in the vector channel 
and $\bar{c} i\gamma_5 D_{\nu}c$ in the axial vector channel. 
}
\end{figure}

The perturbative contribution to $\Pi(Q^2)$ contains a nonlocal divergence 
eliminated through operator mixing under renormalization of the hybrid meson
current~(\ref{hybridCurrent}).
The replacements,
\begin{gather*}
  j^{(h)}_{\nu} \rightarrow j^{(h)}_{\nu} 
  - \frac{5g_s^2 m^2}{18\pi^2\epsilon}\,\overline{c}\gamma_{\nu} c
  + \frac{g_s^2 m}{9\pi^2\epsilon}\,\overline{c} iD_{\nu} c\ \text{for}\ 1^{--},\\ 
  j^{(h)}_{\nu} \rightarrow j^{(h)}_{\nu} 
  - \frac{5g_s^2 m^2}{18\pi^2\epsilon}\,\overline{c}\gamma_{\nu} \gamma_5 c
  - \frac{g_s^2 m}{9\pi^2\epsilon}\,\overline{c} i\gamma_5 D_{\nu}c\ \text{for}\ 1^{++},\\ 
\end{gather*}
for covariant derivative operator $D_{\nu}$
lead to two renormalization-induced diagrams shown in 
Fig.~\ref{renormalizationDiagrams}.
These two diagrams cancel the nonlocal divergence and provide
nontrivial contributions to the finite part of perturbation theory.
Again, calculational details and results can be found in~\cite{Palameta_I,Palameta_II}.

\section{Laplace Sum-Rules}\label{III}
The function $\Pi(Q^2)$ satisfies a dispersion relation,
\begin{equation}\label{dispersion_relation}
  \Pi(Q^2)=\frac{Q^6}{\pi}\int_{t_0}^{\infty}
  \frac{\mathrm{Im}\Pi(t)}{t^3(t+Q^2)}
  \mathrm{d}t +\cdots,
\end{equation}
for $Q^2=-q^2>0$
where $t_0$ is a hadron production threshold.
On the left-hand side, $\Pi(Q^2)$ is identified with
the correlator computed in Section~\ref{II}, denoted $\Pi^{\qcd}(Q^2)$ from here on.
On the right-hand side, the hadronic spectral function, $\mathrm{Im}\Pi(t)$,
is decomposed as
\begin{equation}\label{decomposition}
\frac{1}{\pi} \mathrm{Im}\Pi(t) = \rho^{\text{had}}(t) 
+ \frac{1}{\pi} \mathrm{Im}\Pi^{\qcd}(t)
\Theta(t-s_0)
\end{equation}
where $\Theta(t-s_0)$ is a Heaviside step function at continuum threshold $s_0$
and $\rho^{\had}(t)$, the resonance content, is modelled as 
\begin{equation}\label{hadronContent}
  \rho^{\had}(t) = \sum_{i=1}^n \xi_i \delta(t-m_i^2).
\end{equation}
where $m_i$ are resonance masses and $\xi_i$ are mixing parameters.
A resonance with nonzero mixing parameter couples to both conventional and hybrid meson currents.
Specific resonance models are defined in Tables~\ref{vectorModels} and~\ref{axialModels}
for the vector and axial vector channels respectively.  Note that, in the vector channel,
some densely packed resonances are amalgamated as resonances clusters.
For these clusters, the corresponding
$\delta$-function in~(\ref{hadronContent}) is replaced by a rectangular ``pulse''
to account for the nonzero width $\Gamma$.

Subtracted LSRs are defined as~\cite{SVZ_I,SVZ_II}
\begin{multline}\label{lsrQCD}
  \lsr^{\qcd}\double{\tau}{s_0} =
  \frac{1}{\tau}\!\lim_{\stackrel{N,Q^2\rightarrow\infty}{\tau=N/Q^2}}\!
  \frac{(-Q^2)^N}{\Gamma(N)}
  \left(\frac{\mathrm{d}}{\mathrm{d}Q^2}\right)^N\!\Pi^{\qcd}(Q^2)\\
  -\int_{s_0}^{\infty}\! e^{-t\tau} \frac{1}{\pi}\mathrm{Im}\Pi^{\qcd}(t)\,\mathrm{d}t
\end{multline}
where $\tau$ is the Borel parameter.
Then, eqns.~(\ref{dispersion_relation})--(\ref{lsrQCD}) imply that 
$\lsr^{\qcd}\double{\tau}{s_0}=\lsr^{\had}(\tau;\,\{\xi_i\})$
where~\cite{Palameta_I,Palameta_II}
\begin{equation}\label{lsrHadron}
\lsr^{\had}(\tau;\,\{\xi_i\}) = \int_0^{s_0}\! e^{-t\tau}\rho^{\text{had}}(t)\,\mathrm{d}t.
\end{equation}

\begin{table}
\centering
\caption{A representative collection of hadron models analyzed in the vector sector.}
\label{vectorModels}
\begin{tabular}{|c|c|c|c|c|c|c|}
  \hline
  Model & $m_1$ & $\Gamma_1$ & $m_2$ & $\Gamma_2$ & $m_3$ & $\Gamma_3$\\
  \stud & (GeV) & (GeV) & (GeV) & (GeV) & (GeV) & (GeV)\\ 
  \hline
  V1 & 3.10 & 0 & - & - & - & - \\
  V2 & 3.10 & 0 & 3.73 & 0  & - & - \\
  V3 & 3.10 & 0 & 3.73 & 0 & 4.30 & 0 \\
  V4 & 3.10 & 0 & 3.73 & 0 & 4.30 & 0.30 \\
  V5 & 3.10 & 0 & 3.73 & 0.05 & 4.30 & 0.30 \\
  V6 & 3.10 & 0 & - & - & 4.30 & 0 \\
  V7 & 3.10 & 0 & - & - & 4.30 & 0.30 \\
  \hline
\end{tabular}
\end{table}

\begin{table}
\centering
\caption{A representative collection of hadron models analyzed in the axial vector sector.}
\label{axialModels}
\begin{tabular}{|c|c|c|c|c|}
  \hline
  Model & $m_1$ & $m_2$ & $m_3$ & $m_4$\\
  \stud & (GeV) & (GeV) & (GeV) & (GeV)\\ 
  \hline
  A1 & 3.51 & - & - & -  \\
  A2 & 3.51 & 3.87 & - & -  \\
  A3 & 3.51 & 3.87 & 4.15 & -  \\
  A4 & 3.51 & 3.87 & 4.15 & 4.27  \\
  \hline
\end{tabular}
\end{table}

\section{Analysis and Results}\label{IV}
\begin{figure}[hbt]
\centering
\includegraphics[width=0.9\columnwidth]{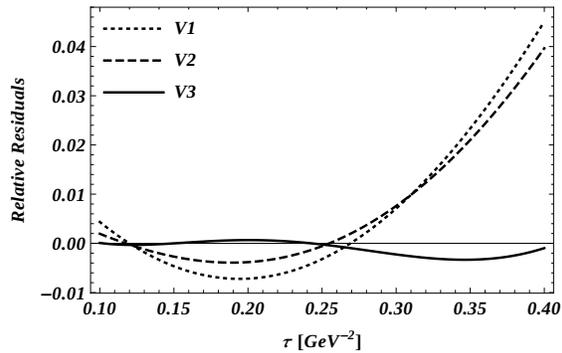}
\caption{\label{vectorResiduals} Relative residuals~(\ref{residuals}) versus Borel 
parameter $\tau$ for a representative set of hadron models in the vector channel 
(\ie\ from Table~\ref{vectorModels}).} 
\end{figure}

\begin{figure}[hbt]
\centering
\includegraphics[width=0.9\columnwidth]{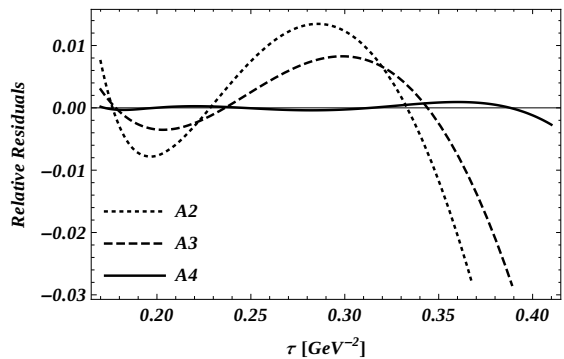}
\caption{\label{axialResiduals} The same as Fig.~\ref{vectorResiduals} 
but for models in the axial vector channel (\ie\ from Table~\ref{axialModels}).} 
\end{figure}

For each of the hadron models of Tables~\ref{vectorModels} and~\ref{axialModels},
we extract mixing parameters $\{\xi_i\}$ and a continuum threshold $s_0$
as best-fit values between~(\ref{lsrQCD}) and~(\ref{lsrHadron}).
To do so, we minimize the chi-square,
\begin{equation}\label{chisq}
  \chi^2(s_0;\,\{\xi_i\}) = \sum^{\tau_{\text{max}}}_{\tau_{\text{min}}}
  \big(\lsr^{\qcd}\double{\tau}{s_0}-\lsr^{\had}(\tau;\,\{\xi_i\})\big)^2,
\end{equation}
over a (discretized) interval of acceptable $\tau$-values $\double{\tau_{\text{min}}}{\tau_\text{max}}$.
(See~\cite{Palameta_I,Palameta_II} for more detail.)
Results are given in Table~\ref{vectorResults} and Table~\ref{axialResults} for the vector and axial vector models respectively.
Instead of $\xi_i$, we present $\zeta$ and $\frac{\xi_i}{\zeta}$ where
\begin{equation}\label{zeta}
  \zeta = \sum_{i=1}^{n} |\xi_i|
\end{equation} 
and where $n$ is the number of resonances in the model in question.
Also, the given minimized values of~(\ref{chisq}) have been scaled by the minimized value for the
single narrow resonance model in each channel, \ie\ Model V1 in the vector channel and Model A1
in the axial vector channel.
We plot relative residuals,
\begin{equation}\label{residuals}
\frac{\lsr^{\qcd}\double{\tau}{s_0}-\lsr^{\had}(\tau,\,\{\xi_i\})}{\lsr^{\qcd}\double{\tau}{s_0}},
\end{equation}
versus $\tau$ in Fig.~\ref{vectorResiduals} for a representative set of vector models and 
in Fig.~\ref{axialResiduals} and for a representative set of axial vector models.

\begin{table*}[hbt]
\begin{center}
\caption{Predicted mixing parameters with theoretical uncertainties 
and continuum thresholds for the vector hadron models of Table~\ref{vectorModels}.}
\label{vectorResults}
\begin{tabular}{|c|c|c|c|c|c|c|}
\hline
    \rule[-1.5ex]{0ex}{4.25ex}Model 
    & $s_0\ (\text{GeV}^2)$
    & $\frac{\chi^2}{\chi^2(V1)}$ 
    & $\zeta\ (\text{GeV}^6)$ 
    & $\frac{\xi_1}{\zeta}$ 
    & $\frac{\xi_2}{\zeta}$ 
    & $\frac{\xi_3}{\zeta}$\\
\hline
  V1 & 12.5 & 1     & 0.51(2) & 1 & {-} & {-} \\
  V2 & 13.9 & 0.73  & 0.73(4) & 0.73(3) &  0.27(3)  & {-}\\
  V3 & 24.1 & 0.038 & 2.9(3)  & 0.22(1) & -0.022(5) & 0.76(3) \\
  V4 & 24.2 & 0.037 & 3.0(3)  & 0.21(1) & -0.032(5) & 0.76(3) \\
  V5 & 24.2 & 0.037 & 3.0(3)  & 0.21(1) & -0.032(5) & 0.76(3) \\
  V6 & 23.7 & 0.042 & 2.7(2)  & 0.23(2) & {-}       & 0.77(2) \\
  V7 & 23.6 & 0.047 & 2.7(2)  & 0.23(2) & {-}       & 0.77(2) \\
\hline
\end{tabular}
\end{center}
\end{table*}

\begin{table*}[hbt]
\begin{center}
\caption{The same as Table~\ref{vectorResults} but for the hadron models of Table~\ref{axialModels}.}
\label{axialResults}
\begin{tabular}{|c|c|c|c|c|c|c|c|}
\hline
    \rule[-1.5ex]{0ex}{4.25ex}Model 
    & $s_0\ (\text{GeV}^2)$
    & $\frac{\chi^2}{\chi^2(V1)}$ 
    & $\zeta\ (\text{GeV}^6)$ 
    & $\frac{\xi_1}{\zeta}$ 
    & $\frac{\xi_2}{\zeta}$ 
    & $\frac{\xi_3}{\zeta}$
    & $\frac{\xi_4}{\zeta}$\\
\hline
  A1 & 18.8 & 1       & 0.18(1) & 1 & {-} & {-} & {-} \\
  A2 & 28.8 & 0.0095  & 0.83(7) & 0.47(2) & -0.53(2) & {-} & {-}\\
  A3 & 18.8 & 0.0034  & 2.6(4)  & 0.21(2) & -0.45(1) & 0.34(2) & {-} \\
  A4 & 31.7 & $7.3\times10^{-6}$  & 44(6)   & 0.03(1) & -0.16(1) & 0.46(1) & -0.35(1) \\
\hline
\end{tabular}
\end{center}
\end{table*}

\section{Discussion}\label{V}
From the normalized chi-squares in Tables~\ref{vectorResults} and~\ref{axialResults},
we see that agreement between theoretically calculated LSRs and hadron physics 
is significantly improved by including both excited resonances and the ground 
state in the hadron model.  Based on chi-squares values 
and the relative residuals plotted in Figs.~\ref{vectorResiduals} and~\ref{axialResiduals}, 
we favour Models V3--V7 in the vector channel (all of which lead to essentially the same conclusions)
and Model A4 in the axial vector channel.  
Note that the resonance widths in Models V4, V5, and~V7 have little effect on the results. 
This is unsurprising as LSRs are generally insensitive to symmetric resonance widths.
By design, LSRs exponentially suppress contributions 
from heavy resonances relative to lighter ones, and so it is important 
to check that the heavy resonances of Models V3--V7 and A4 
make numerically significant contributions. 
As a quantitative measure in, for example, Model V3, consider
\begin{equation}
\frac%
{|\xi_3| \int_{\tau_{\text{min}}}^{\tau_\text{max}}\! e^{-m_3^2 \tau}\,\mathrm{d}\tau}%
{\sum_{i=1}^3 |\xi_i|  \int_{\tau_{\text{min}}}^{\tau_\text{max}}\! e^{-m_i^2 \tau}\,\mathrm{d}\tau},
\end{equation}
\ie\ the ratio of the heaviest resonance's contribution to the LSRs
to the total resonance contribution to the sum-rules.  Using values of $m_i$
and $\xi_i$ from Tables~\ref{vectorModels} and~\ref{vectorResults} respectively, 
this ratio evaluates to 0.43.  
In the axial vector channel, an analogous ratio measuring the relative contribution to 
the LSRs of $m_4$ gives 0.25.

Employing QCD LSRs, we studied conventionl meson-hybrid meson mixing in vector and axial vector charmonium-like channels.  
Using measured masses as inputs, we tested experimentally observed resonances for coupling 
to both conventional and hybrid meson currents, \ie\ for meson-hybrid mixing.  In both channels,
agreement between QCD and hadron physics was significantly improved by the inclusion 
of resonances above 4 GeV.  In the vector channel, we found that conventional meson-hybrid meson mixing was 
well-described by a two resonance scenario consisting of the $J/\psi$ and a 4.3~GeV state.  
These results are consistent with the $J/\psi$ being predominantly a conventional meson but with a small hybrid meson component.  
As for the heavier state, it has been speculated that the $Y(4260)$ 
has a significant hybrid meson component (see~\cite{Zhu}, for example), an interpretation 
consistent with our findings.  In the axial vector channel, we found almost no mixing in the 
ground state, $\chi_{c1}(1P)$, minimal mixing in the $X(3872)$, and significant mixing in both 
the $X(4140)$ and $X(4274)$.  Ref.~\cite{Matheus} argues that the $X(3872)$ has a significant conventional meson 
component while Ref.~\cite{Chen} argues that it has a significant hybrid meson component.  
Our results are compatible with either conclusion, but have difficulty accommodating both.  

\begin{acknowledgments}
We are grateful for financial support from the National Sciences and Engineering Research Council of Canada.
\end{acknowledgments}

\bigskip 

\begin{thebibliography}{9}   

\bibitem{Brambilla} N.\ Brambilla \textit{et al.,} Eur.\ Phys.\ J.\ \textbf{C71} (2011) 1534, 1010.5827.
\bibitem{PDG} Particle Data Group, C.\ Patrignani \textit{et al.,} Chin. Phys.\ \textbf{C40} (2019) 100001.
\bibitem{SVZ_I} M.\ A.\ Shifman, A.\ I.\ Vainshtein, and V.\ I.\ Zakharov, Nucl.\ Phys.\ \textbf{B147} (1979) 385.
\bibitem{SVZ_II} M.\ A.\ Shifman, A.\ I.\ Vainshtein, and V.\ I.\ Zakharov, Nucl.\ Phys.\ \textbf{B147} (1979) 448.
\bibitem{Palameta_I} A.\ Palameta, J.\ Ho, D.\ Harnett, and T.\ G.\ Steele, Phys.\ Rev.\ \textbf{D97} (2018) 034001, 1707.00063.
\bibitem{Palameta_II} A.\ Palameta, D.\ Harnett, and T.\ G.\ Steele, Phys.\ Rev.\ D98 (2018) 074014, 1805.04230.
\bibitem{jaxodraw} D.\ Binosi \textit{et al.,}, Compute.\ Phys.\ Commun.\ \textbf{180} (2009) 1709, 0811.4113.
\bibitem{Zhu} S.-L.\ Zhu, Phys.\ Lett.\ \textbf{B631} (2005) 212, hep-ph/0507025.
\bibitem{Matheus} R.\ D.\ Matheus \emph{et al.,} Phys.\ Rev.\ \textbf{D80} (2009) 056002, 0907.2683.
\bibitem{Chen} W.\ Chen \textit{et al.,} Phys.\ Rev.\ \textbf{D88} (2013) 045027, 1305.0244.


\end{thebibliography}

\end{document}